\documentclass[12pt]{article} 

\usepackage{amsmath}  
\usepackage{amsfonts} 
\usepackage{graphicx} 
\usepackage{url}
\usepackage[bottom]{footmisc}
\usepackage{amssymb}
\usepackage{booktabs}
\usepackage[labelformat=simple]{subcaption}
\usepackage{textcomp}
\usepackage{mathrsfs}
\usepackage[shortlabels]{enumitem}
\usepackage{upgreek}
\usepackage{fullpage}
\usepackage[title]{appendix}
\usepackage[margin=1in]{geometry}
\usepackage{color,soul}
\usepackage{times}
\usepackage{pgf,tikz}
\usepackage[raggedright]{titlesec}
\usepackage{multirow}
\usepackage{multicol}
\usepackage{array}
\usepackage{setspace}
\usepackage[format=plain,labelsep=period]{caption}
\usepackage{cite}
\usepackage{esint}
\usepackage{listings}
\usepackage{fancyhdr}

\usepackage{hyperref}
\hypersetup{colorlinks=true,linkcolor=blue} 

\titleformat{\section}{\bfseries}{\thesection}{1em}{}
\titleformat{\subsection}{\itshape}{\thesubsection}{1em}{}
\titleformat{\subsubsection}{}{\thesubsubsection}{1em}{}

\newcolumntype{L}[1]{>{\raggedright\let\newline\\\arraybackslash\hspace{0pt}}m{#1}}
\newcolumntype{C}[1]{>{\centering\let\newline\\\arraybackslash\hspace{0pt}}m{#1}}
\newcolumntype{R}[1]{>{\raggedleft\let\newline\\\arraybackslash\hspace{0pt}}m{#1}}

\begin{document}

\begin{table}[t]
\centering
\begin{tabular}{c}
 {\large \textbf{A canonical Hamiltonian formulation of the Navier-Stokes problem}} \\ \\
\end{tabular}
\begin{tabular}{ccc}
John W. Sanders$^{*,a}$ & A. C. DeVoria$^{a}$ & Nathan J. Washuta$^{a}$ \\ Gafar A. Elamin$^{a}$ & Kevin L. Skenes$^{a}$ & Joel C. Berlinghieri$^{b}$ \\ \\
\end{tabular}
\begin{tabular}{c}
{$^{*}$Corresponding Author: jsande12@citadel.edu} \\ \\
{$^{a}$Department of Mechanical Engineering} \\
{The Citadel, the Military College of South Carolina} \\
{171 Moultrie St, Charleston, SC 29409} \\ \\
{$^{b}$Department of Physics} \\
{The Citadel, the Military College of South Carolina} \\
{171 Moultrie St, Charleston, SC 29409}
\end{tabular}
\end{table}

\section*{Abstract}

This paper presents a novel Hamiltonian formulation of the isotropic Navier-Stokes problem based on a minimum-action principle derived from the principle of least squares. This formulation uses the velocities $u_{i}(x_{j},t)$ and pressure $p(x_{j},t)$ as the field quantities to be varied, along with canonically conjugate momenta deduced from the analysis. From these, a conserved Hamiltonian functional $H^{*}$ satisfying Hamilton's canonical equations is constructed, and the associated Hamilton-Jacobi equation is formulated for both compressible and incompressible flows. This Hamilton-Jacobi equation reduces the problem of finding four separate field quantities ($u_{i}$,$p$) to that of finding a single scalar functional in those fields---Hamilton's principal functional $\text{S}^{*}[u_{i},p,t]$. Moreover, the transformation theory of Hamilton and Jacobi now provides a prescribed recipe for solving the Navier-Stokes problem: Find $\text{S}^{*}$. If an analytical expression for $\text{S}^{*}$ can be obtained, it will lead via canonical transformation to a new set of fields which are simply equal to their initial values, giving analytical expressions for the original velocity and pressure fields. Failing that, if one can only show that a complete solution to this Hamilton-Jacobi equation does or does not exist, that will also resolve the question of existence of solutions. The method employed here is not specific to the Navier-Stokes problem or even to classical mechanics, and can be applied to any traditionally non-Hamiltonian problem. 

\section{Introduction}\label{sec:intro}

Given the title of this paper, it is incumbent on the authors to assure the reader that we do not claim to have done the impossible. A viscous fluid is, after all, a non-Hamiltonian system~\cite{Millikan1929,Finlayson1972,Finlayson1972a}. There is no action integral for which Hamilton's principle~\cite{Hamilton1833,Hamilton1834,Hamilton1835} yields the Navier-Stokes equations~\cite{Stokes1845,Anderson1984,Anderson1995,Batchelor2000,White2006,Pozrikidis2009,Cengel2018} in their usual form~\cite{Millikan1929,Finlayson1972,Finlayson1972a}, and we do not claim otherwise. Remarkably, however, a Hamiltonian formulation can still be found by considering a \emph{mathematically equivalent higher-order problem}, as we will now demonstrate via simple example.

\subsection{A motivating example}\label{sec:toyproblem}

Consider the first-order initial-value problem
\begin{equation}\label{eq:toy1}
\dot{v}=-v, \quad v(0)=1,
\end{equation}
with unique solution $v(t)=e^{-t}$. Here $v(t)$ can be interpreted as the velocity of a lumped mass moving in a viscous medium in one dimension with linear damping. Like the traditional Navier-Stokes equations~\cite{Stokes1845,Anderson1984,Anderson1995,Batchelor2000,White2006,Pozrikidis2009,Cengel2018}, this too is an intrinsically non-Hamiltonian problem, in that there is no action $\mathcal{S}$ for which Hamilton's principle ($\delta\mathcal{S}=0$) yields the governing equation $\dot{v}=-v$. Yet if we simply differentiate both sides of the equation ($\ddot{v}=-\dot{v}$), use the original equation to write $\dot{v}=-v$, and apply the additional initial condition $\dot{v}(0)=-v(0)=-1$, we arrive at the mathematically equivalent second-order problem
\begin{equation}\label{eq:toy2}
\ddot{v}=v, \quad v(0)=1, \quad \dot{v}(0)=-1,
\end{equation}
which has the same unique solution $v(t)=e^{-t}$ but which \emph{is} Hamiltonian---not in the sense that the total mechanical energy is conserved, but in the sense that it has mathematically Hamiltonian structure. 

As first observed by Sanders~\cite{Sanders2021,Sanders2022,Sanders2023a,Sanders2023,Sanders2023b}, the associated action can be obtained by writing the original equation in standard form ($\mathcal{R}\equiv\dot{v}+v=0$), squaring the residual $\mathcal{R}$, and integrating over time:
\begin{equation}\label{eq:toyaction}
\mathcal{S}^{*}[v]=\int\text{d}t\left(\frac{1}{2}\mathcal{R}^{2}\right)=\int\text{d}t\left[\frac{1}{2}\left(\dot{v}^{2}+2v\dot{v}+v^{2}\right)\right]\sim\int\text{d}t\left[\frac{1}{2}\left(\dot{v}^{2}+v^{2}\right)\right],
\end{equation}
where we have used the fact that $2v\dot{v}=\text{d}(v^{2})/\text{d}t$ is a total time derivative and can therefore be excluded from the action without changing the resulting Euler-Lagrange equation~\cite{Lanczos1986}. This is the so-called ``time-averaged principle of least squares''~\cite{Sanders2021,Sanders2022,Sanders2023a,Sanders2023,Sanders2023b}: since $\mathcal{R}=0$ is a local minimum of $\mathcal{R}^{2}$, it is also a local minimum of $\int \text{d}t(\mathcal{R}^{2})$. Varying $v$, the first variation of $\mathcal{S}^{*}$ is
\begin{equation}
\delta\mathcal{S}^{*}=\int\text{d}t\biggl[\dot{v}\delta\dot{v}+v\delta v\biggr]=\int\text{d}t\biggl[(-\ddot{v}+v)\delta v\biggr]+\biggl[\dot{v}\delta v\biggr]_{t_{1}}^{t_{2}},
\end{equation}
yielding the second-order equation $\ddot{v}=v$ and revealing the canonically conjugate ``momentum'' $\pi\equiv\dot{v}$. Here and in what follows, we will use the symbol $\pi$ for canonically conjugate momenta, as is customary in Hamiltonian field theory, in order to avoid later confusion with the pressure field $p$. Since the mathematical constant $3.14159...$ does not appear in the present work, there will be no ambiguity.

The corresponding Hamiltonian is obtained via the Legendre transform
\begin{equation}
H^{*}[v,\pi]=\pi\dot{v}-\frac{1}{2}\left(\dot{v}^{2}+v^{2}\right)=\frac{1}{2}\left(\pi^{2}-v^{2}\right).
\end{equation}
Notably, this Hamiltonian has nothing to do with the total mechanical energy of the system, although it \emph{is} a conserved quantity. In fact, $H^{*}=0$ for the actual motion satisfying $\pi\equiv\dot{v}=-v$. We note in passing that Liouville's theorem is satisfied, as the motion occurs along the line $\pi=-v$, so that the phase-space volume, being always zero, is conserved. Hamilton's equations
\begin{equation}
\dot{v}=\frac{\partial H^{*}}{\partial\pi}, \quad 
\dot{\pi}=-\frac{\partial H^{*}}{\partial v}
\end{equation}
are mathematically equivalent to the second-order problem $\ddot{v}=v$ and therefore also mathematically equivalent to the original, first-order problem.

The associated Hamilton-Jacobi equation~\cite{Hamilton1833,Hamilton1834,Hamilton1835,Jacobi1837,Jacobi18421843,Whittaker,Lanczos1986} is
\begin{equation}
\frac{1}{2}\left(\frac{\partial\text{S}^{*}}{\partial v}\right)^{2}-\frac{1}{2}v^{2}+\frac{\partial\text{S}^{*}}{\partial t}=0,
\end{equation}
where Hamilton's principal function $\text{S}^{*}=\text{S}^{*}(v,t)$ serves as the generating function for a canonical transformation to a new coordinate $\phi$ which is constant and equal to its initial value. Although this is almost identical to the Hamilton-Jacobi equation for the simple harmonic oscillator---the only difference being the sign in front of $(1/2)v^{2}$---the usual separable solution of the form $\text{S}^{*}(v,t)=W(v)+T(t)$ does not work, as the reader may check.

Instead, let us use a trial solution of the form
\begin{equation}
\text{S}^{*}(v,t)=F(t)v+\frac{1}{2}v^{2}+f(t),
\end{equation}
where $F(t)$ and $f(t)$ are as yet undetermined functions of $t$. This trial solution was chosen to cancel the term $(1/2)v^{2}$ from the equation. Substituting our trial solution into the Hamilton-Jacobi equation, we find that
\begin{equation}
\frac{1}{2}[F(t)]^{2}+[F(t)+F'(t)]v+f'(t)=0.
\end{equation}
In order for this equation to hold for all $v$, we must have the following:
\begin{equation}
F(t)+F'(t)=0 \quad \Rightarrow \quad F(t)=\alpha e^{-t},
\end{equation}
where $\alpha$ is a constant of integration which will be used to transform to the new coordinate, and
\begin{equation}
\frac{1}{2}[F(t)]^{2}+f'(t)=0 \quad \Rightarrow \quad f(t)=\frac{1}{4}\alpha^{2}e^{-2t}+\gamma,
\end{equation}
where $\gamma$ is another constant of integration which is simply additive and can therefore be discarded.

In this way, we have that
\begin{equation}
\text{S}^{*}(v,t;\alpha)=\alpha e^{-t}v+\frac{1}{2}v^{2}+\frac{1}{4}\alpha^{2}e^{-2t}.
\end{equation}
With one constant of integration ($\alpha$) to match the single degree of freedom ($v$), this is a complete solution to the Hamilton-Jacobi equation. The new coordinate $\phi$ (which is constant and equal to its initial value) is obtained via the canonical transformation
\begin{equation}
\phi=\frac{\partial\text{S}^{*}}{\partial\alpha}=e^{-t}v+\frac{1}{2}\alpha e^{-2t}.
\end{equation}
The numerical value of $\alpha$ is in turn obtained via the canonical relation
\begin{equation}
\pi=\frac{\partial\text{S}^{*}}{\partial v}=\alpha e^{-t}+v,
\end{equation}
which, evaluated at $t=0$, gives $\alpha=-2$ (recall that $\pi=\dot{v}$, and $\dot{v}(0)=-v(0)=-1$). Using the fact that the new coordinate $\phi$ is equal to its initial value, we have that
\begin{equation}
e^{-t}v-e^{-2t}=v(0)-1=0,
\end{equation}
giving the correct solution $v(t)=e^{-t}$. 

In summary, by doubling the order of the governing equation and supplying additional auxiliary conditions, we made a non-Hamiltonian problem into a Hamiltonian one~\cite{Sanders2021,Sanders2022,Sanders2023a,Sanders2023,Sanders2023b}. Furthermore, this simple example demonstrates that the method correctly gives the solution to the original, non-Hamiltonian problem. Indeed, \emph{it would appear that every non-Hamiltonian problem belongs to an equivalence class of problems with the same solution, and within each such equivalence class there are Hamiltonian variants}. The remainder of this paper applies that concept to the isotropic Navier-Stokes problem~\cite{Stokes1845,Anderson1984,Anderson1995,Batchelor2000,White2006,Pozrikidis2009,Cengel2018}.

\subsection{The Navier-Stokes problem}

The incompressible Navier-Stokes equations~\cite{Stokes1845,Anderson1984,Anderson1995,Batchelor2000,White2006,Pozrikidis2009,Cengel2018} are given by
\begin{equation}\label{eq:NS0}
\rho \dot{u}_{i} + \rho u_{i,j}u_{j} + p_{,i} - \mu u_{i,jj} - \rho b_{i} = 0,
\end{equation}
\begin{equation}\label{eq:continuity0}
u_{i,i} = 0,
\end{equation}
where $\rho$ is the constant and uniform density, $u_{i}=u_{i}(x_{j},t)$ is the velocity field, $p=p(x_{j},t)$ is the pressure field, $b_{i}=b_{i}(x_{j},t)$ is the body force field, subscript Roman indices label Euclidean tensor components ($i,j=1,2,3$), the $x_{j}$ are Eulerian spatial coordinates, $t$ is time, $\mu$ is the dynamic viscosity, a dot over a symbol denotes a \emph{partial} time derivative ($\dot{u}_{i}=\partial{u}_{i}/\partial t$), a comma in a subscript indicates a spatial gradient ($p_{,i}=\partial p/\partial x_{i}$), and we employ the Einstein summation convention on repeated subscript indices. To be clear, the notation $u_{i}(x_{j},t)$ signifies that each component of the velocity field is a function of all three spatial coordinates $(x_{1},x_{2},x_{3})$ and time $t$. It has the same meaning as other common notations, such as $u_{i}(\mathbf{x},t)$ and $\mathbf{u}(\mathbf{x},t)$. Likewise for all other field quantities. In the case of a uniform gravitational field, $b_{i}=g_{i}$ coincides with the local acceleration due to gravity; however, in what follows, we make no assumptions about the functional form of $b_{i}(x_{j},t)$: it is completely arbitrary. There are four unknown field quantities: $u_{i}(x_{j},t)$ and $p(x_{j},t)$.

We seek, ultimately, a functional 
\begin{equation}
H^{*}=H^{*}[u_{i},p,\pi_{j},\pi_{4};t],
\end{equation}
where ($\pi_{i},\pi_{4}$) are suitable ``momenta'' conjugate to the field quantities ($u_{i},p$), such that Hamilton's canonical equations
\begin{align}
\dot{u}_{i}&=\frac{\delta H^{*}}{\delta\pi_{i}}, &\dot{p}&=\frac{\delta H^{*}}{\delta\pi_{4}}, \label{eq:Hamilton120} \\
\dot{\pi}_{i}&=-\frac{\delta H^{*}}{\delta u_{i}}, &\dot{\pi}_{4}&= -\frac{\delta H^{*}}{\delta p}, \label{eq:Hamilton340}
\end{align}
constitute a mathematically equivalent \emph{second-order} formulation of the problem, where $\delta H^{*}/\delta u_{i}$, $\delta H^{*}/\delta p$, $\delta H^{*}/\delta \pi_{i}$, and $\delta H^{*}/\delta \pi_{4}$ are the Volterra~\cite{Volterra1930} functional derivatives of $H^{*}$ with respect to the field quantities and the conjugate momenta. We will find that this is generally possible for a compressible fluid. For an incompressible fluid, the equation $\dot{p}=\delta H^{*}/\delta\pi_{4}$ will need to be replaced by the incompressibility condition $u_{i,i}=0$, consistent with the well known result that the pressure usually serves as a Lagrange multiplier for the incompressibility constraint~\cite{Lanczos1986,Badin2018} (refer to p. 361 of~\cite{Lanczos1986} and pp. 137 and 141 of~\cite{Badin2018}).

The remainder of this paper is organized as follows. Section~\ref{sec:literature} gives a comprehensive overview of the relevant literature to date. Sections~\ref{sec:LagrangianformulationofNSP} and~\ref{sec:HamiltonianformulationofNSP} contain the main results of the present work, culminating in a conserved Hamiltonian functional $H^{*}$ satisfying Hamilton's equations~\eqref{eq:Hamilton120} and~\eqref{eq:Hamilton340} for the mathematically equivalent second-order problem, along with the accompanying Hamilton-Jacobi equation~\cite{Hamilton1833,Hamilton1834,Hamilton1835,Jacobi1837,Jacobi18421843, Whittaker,Lanczos1986}. Section~\ref{sec:discussion} contains a discussion of the physical interpretation of the second-order formulation. Section~\ref{sec:examples} presents a brief case study in the form of one-dimensional flow over an infinite, flat plate. Finally, Section~\ref{sec:conclusion} concludes the paper with a few closing remarks and an outline of how the present formulation can aid in resolving the question of existence and uniqueness of solutions to the Navier-Stokes problem. 

By the end of the paper, we will have achieved precisely what the title promises: a canonical Hamiltonian formulation of the problem, opening new avenues toward resolution of one of the most famous unsolved problems in mathematics.

\section{Literature review}\label{sec:literature}

The field of analytical mechanics, with foundations planted in Hamilton's principle of stationary action~\cite{Hamilton1833,Hamilton1834,Hamilton1835} or d'Alembert's principle of virtual work~\cite{dAlembert1743}, has been vital to the development of both classical and quantum physics since the eighteenth century.  This approach is versatile and helpful to the physical understanding of the problem in question, and the foundation, structure, and utility of Hamiltonian formalism is well-documented~\cite{Becker1954,Taylor2005,Hamill2014,Bohn2018,Cline2023,Fowler2023}. The supporting mathematics of the calculus of variations as well as symplectic and differential geometry can also be found in many excellent sources~\cite{Arnold1989,Berndt2001,Hall2003,Boas2006,Stone2009,Gelfand2012,Arfken2013,Needham2021}. It is therefore no surprise that researchers have been applying analytical formalism to classical fluids dating back to the time of Lagrange~\cite{Lagrange1811,Lichtenstein1929,Morrison1998,Morrison2006,Berdichevsky2009,Berdichevsky2009a,dellIsola2011,Badin2018,Bedford2021}. 

The task of obtaining solutions to the governing equations of fluid flow represents one of the most challenging problems in science and engineering. In most cases, the mathematical formulation is expressed as an initial-boundary-value problem: a set of coupled, nonlinear partial differential equations, which are to be solved subject to various initial- and boundary conditions. The degree of complication of the governing equations depends on the type of the fluid. For a viscous fluid where the transport phenomena of friction and thermal conduction are included, the governing equations are called the Navier-Stokes equations~\cite{Stokes1845,Anderson1984,Anderson1995,Batchelor2000,White2006,Pozrikidis2009,Cengel2018}. The Navier-Stokes equations are derived by applying fundamental physical principles---conservation of mass, conservation of momentum, and conservation of energy---to a viscous fluid, and the derivation can be found in any fluid mechanics textbook~\cite{Anderson1984,Anderson1995,Batchelor2000,White2006,Pozrikidis2009,Cengel2018}. It is important to recognize that the Navier-Stokes equations as they are known today were not developed solely by Navier and Stokes; indeed, Poisson, Cauchy, and others were also heavily involved in their development~\cite{Darrigol2002}. As far as the present authors are aware, to date there is still no firm answer to the question of whether or not there always exist unique, smooth, nonsingular solutions to the three-dimensional Navier-Stokes equations~\cite{Lemarie2018}, and this constitutes one of the most famous unsolved problems in mathematics.

The application of analytical mechanics~\cite{Goldstein1980,Arnold1989,Fetter2003,Gelfand2012} to the field of fluid mechanics~\cite{Lanczos1986} has recently seen a resurgence in interest~\cite{Salmon1983,Salmon1988,Brenier2017,Giga2018,Mottaghi2019,Taroco2020,Bedford2021,Mavroeidis2022} after a long history. In the absence of non-conservative forces, an inviscid fluid is a Hamiltonian system, and so the classical Hamiltonian theory applies. Serrin~\cite{Serrin1959}, Benjamin~\cite{Benjamin1984}, and Holm~\emph{et al.}~\cite{Holm1986} have all described variational and Hamiltonian formulations of incompressible, inviscid fluid flow.  Roberts~\cite{Roberts1972} presented a Hamiltonian dynamic for weakly interacting vortices. This research obtained the canonical equations of Hamiltonian dynamics for a set of two well-separated vortex rings by setting up a Hamiltonian to define the set. Olver~\cite{Olver1982} showed that the Euler equations of inviscid and incompressible fluid flow can be put into Hamiltonian form. Benjamin and Olver~\cite{Benjamin1982} investigated the Hamiltonian structure of the water waves problem. They examined the symmetry groups of this problem, finding that Hamiltonian analysis enables the solution of conservative elements of the problem. However, the study also acknowledged that further study is needed to identify the physical significance of the mathematical results. Maddocks and Pego~\cite{Maddocks1995} presented a novel Hamiltonian formulation of ideal fluid flow expressed in material coordinates. Their Hamiltonian formulation arises from a general approach for constrained systems that is not restricted to problems in fluid mechanics. Rather, it is widely applicable for obtaining unconstrained Hamiltonian dynamical systems from Lagrangian field equations that are subject to pointwise constraints. More recently, Arnold~\cite{Arnold2014} also studied the Hamiltonian nature of the ideal Euler equations. 

Viscous forces are non-conservative, which presents a fundamental challenge when applying Hamilton's principle to viscous fluids~\cite{Millikan1929,Finlayson1972,Finlayson1972a,Lemarie2018}. Indeed, it is a well known theorem (first proven by Millikan~\cite{Millikan1929}) that the Navier-Stokes equations in their usual form cannot be derived from a classical action principle~\cite{Millikan1929,Finlayson1972,Finlayson1972a}. Millikan~\cite{Millikan1929} summarizes his main result as follows:
\begin{quote}
It is impossible to derive the equations of steady motion of a viscous, incompressible fluid from a variation principle involving as Lagrangian function an expression in the velocity components and their first-order space derivatives, unless conditions are imposed on these velocity components such that all of the terms $vu_{,2}$, $wu_{,3}$, $wv_{,3}$, $uv_{,1}$, $uw_{,1}$, $vw_{,2}$ disappear from their positions in the Navier-Stokes equations~\cite{Millikan1929}.
\end{quote}
(It should be noted that the six terms referred to above come from the convective acceleration $u_{i,j}u_{j}$, and Millikan~\cite{Millikan1929} uses the notation $u=u_{1}$, $v=u_{2}$, and $w=u_{3}$.) More generally, it has been shown that the existence of variational formulations is related to self-adjointness of the system with respect to a standard duality relation, a property that all non-conservative systems lack~\cite{Vainberg1964}. Within the last 80 years, many alternative methods have been developed in an attempt to circumvent the non-self-adjointness of dissipative systems~\cite{Prigogine1965,Biot1970,Finlayson1972,Lebon1973,Tonti1973,Magri1974,Telega1979,Tonti1984,Filippov1989,Sieniutycz2000,Robinson2001,Galley2013,Kim2016,Mottaghi2019,Taroco2020,Junker2021,Bersani2021}. The mathematical study of alternative variational methods as applied to the Navier-Stokes equations in particular remains an ongoing endeavor~\cite{Oseledets1989,Vujanovic1989,Doering1995,Fukagawa2012,Jones2015,Hieber2017,GayBalmaz2017,Hochgerner2018,GayBalmaz2019,GayBalmaz2019a,Rashad2021,Gonzalez2022,Taha2022,Sanders2023b}.

Oseledets~\cite{Oseledets1989} attempted to express the Navier-Stokes equations using Hamiltonian formalism. He was able to formalize the incompressible Euler equation but stated that his result is not valid for a compressible fluid. More recent attempts, such as Fukagawa and Fujitani~\cite{Fukagawa2012}, Jones~\cite{Jones2015}, and Gay-Balmaz and Yoshimura~\cite{GayBalmaz2017,GayBalmaz2019,GayBalmaz2019a}, have enforced dissipation using a non-holonomic constraint on the entropy. Hochgerner~\cite{Hochgerner2018} attempted to obtain a Hamiltonian interacting particle system that could accurately model fluid dynamics. His research separated the dynamics into slow (deterministic) and fast (stochastic) components to capture fine-scale effects. The study was able to derive the Navier-Stokes equation from a stochastic Hamiltonian system but ignored the stress tensor, was unable to separate configuration and momentum variables, and did not establish energy conservation or dissipation.

Rashad \emph{et al.}~\cite{Rashad2021} modeled the incompressible Navier-Stokes equations in so-called ``port-Hamiltonian'' framework rather than the standard Hamiltonian framework. Their model used vector calculus instead of exterior calculus to minimize the number of operators. While the main goal of this research was increasing the interest of computational researchers in using vector calculus, they also demonstrated that vector calculus can help in the formulation of individual subsystems of Navier-Stokes equations and boundary ports of the model.

Gonzalez and Taha~\cite{Gonzalez2022}, Taha and Gonzalez~\cite{Taha2022}, and Taha, Gonzalez, and Shorbagy~\cite{Taha2023} have recently applied Gauss's principle of least constraint~\cite{Gauss1829} to the Navier-Stokes problem. Using Gauss's principle, Taha, Gonzalez, and Shorbagy~\cite{Taha2023} have shown that, for an incompressible fluid, the magnitude of the pressure gradient is minimum over the domain, which they term the Principle of Minimum Pressure Gradient (PMPG). When applied to an inviscid fluid in two dimensions, the PMPG provides a closure condition based in first principles that could be considered a generalization of the Kutta condition to smooth geometries. It should be noted that Gauss's principle~\cite{Gauss1829} is fundamentally different from Hamilton's principle~\cite{Hamilton1833,Hamilton1834,Hamilton1835}. Whereas the Hamiltonian framework involves an invariant action integral and employs variations in the coordinates (or, in continuum mechanics, the field quantities), Gauss's principle employs variations in the accelerations. As a result, the framework of Gauss's principle does not lead to canonical transformations.

Particularly relevant to the present work, Sanders~\cite{Sanders2021,Sanders2022,Sanders2023a,Sanders2023,Sanders2023b} has shown that higher-order dynamics are ``intrinsically variational,'' in the sense that higher-derivative versions of the classical equations of motion can be derived from a minimum action principle even for dissipative systems, thus allowing inherently non-Hamiltonian problems to be treated as though they are Hamiltonian. This discovery has already led to two applications: the direct modal analysis of damped dynamical systems~\cite{Sanders2022} and subsequently a new and more efficient algorithm for computing a damped system's resonant frequencies~\cite{Sanders2023}. Higher-derivative theories had been studied before in the realm of quantum gravity physics~\cite{Pais1950,VandenBerg2002,Kalies2004,Bender2008,Smilga2009,Mostafazadeh2010,Baleanu2012,Chen2013} but until now they have not been applied to classical fluids. While the Navier-Stokes equations, in their standard form, may be unsuited to Hamiltonian formalism~\cite{Millikan1929,Finlayson1972,Finlayson1972a,Doering1995,Hieber2017,Lemarie2018}, it will be shown here that higher-order dynamics can be used to restate the problem in a form consistent with Hamiltonian and Hamilton-Jacobi formalism.  

In conclusion, although the body of research surrounding the Navier-Stokes equations is extensive, it would appear that no canonical Hamiltonian formulation of the Navier-Stokes problem has been found to date. That is what the present work aims to achieve.

\section{Lagrangian formulation of the problem}\label{sec:LagrangianformulationofNSP}

Although we are primarily interested in the incompressible form of the equations given by~\eqref{eq:NS0} and \eqref{eq:continuity0}, here we will begin with the compressible form of the equations, with the understanding that we will eventually take the incompressible limit. For the compressible case, the linear momentum balance and continuity equations are given by
\begin{equation}\label{eq:NS1}
\mathcal{R}_{i}[u_{j},p,\rho;x_{k},t]\equiv \rho \dot{u}_{i} + \rho u_{i,j}u_{j} + p_{,i} - \mu u_{i,jj} - (\mu+\lambda)u_{j,ji} - \rho b_{i} = 0,
\end{equation}
\begin{equation}\label{eq:continuity1}
\mathcal{R}_{4}[u_{j},\rho]\equiv \dot{\rho} + \rho_{,i}u_{i} + \rho u_{i,i} = 0,
\end{equation}
where $\rho=\rho(x_{j},t)$ is the density field (now one of the unknown field quantities along with $u_{i}$ and $p$), and $\lambda$ is an additional viscosity coefficient which, under Stokes's~\cite{Stokes1845} hypothesis, is related to $\mu$ as $\lambda=-2\mu/3$, ensuring that the mechanical pressure agrees with the thermodynamic pressure. Henceforth we will assume that all quantities have been suitably non-dimensionalized. The non-dimensional (constant) viscosities in \eqref{eq:NS1} and \eqref{eq:continuity1} may be regarded as inverse Reynolds numbers, and the non-dimensional pressure may be considered to be normalized by the inertial scale $\rho_0 U^2$, with $\rho_0$ and $U$ appropriate density and velocity scales. As we will see later, starting from the compressible form of the equations will allow us to treat the pressure as a dynamical field variable alongside the velocities, rather than simply a Lagrange multiplier. Crucially, this will reveal in no uncertain terms what becomes of the momentum conjugate to the pressure (which will be identified later) in the incompressible limit.

In general, \eqref{eq:NS1} and \eqref{eq:continuity1} would be appended with the energy equation, which introduces additional thermodynamic variables, such as temperature and enthalpy or entropy. Two of the thermodynamic variables are designated as ``primary,'' and equations of state are required to relate the remaining variables to the primary variables. Typically, pressure and temperature are chosen as the primary variables, and the equation of state for the density, for example, is expressed as $\rho=\rho(p,T)$. The conservation equations along with the equation of state constitute six equations for the six unknowns fields $(u_i,p,T,\rho)$. Henceforth in the present work, we will take the temperature to be constant, though we intend to consider variations in temperature in future work.

An incompressible flow is one for which the material derivative of the density vanishes, \emph{i.e.}, $\text{d}\rho/\text{d}t=\dot{\rho} + \rho_{,i}u_{i}=0$, and this condition serves as an equation of state. It is usually also assumed, for the sake of simplicity, that the density is both constant and uniform, further reducing the equation of state $\rho=\rho(p,T)$ to specification of $\rho=\rho_0$ as a system parameter. Consequently, \eqref{eq:continuity1} reduces to $\rho u_{i,i}=0$ and the energy equation is decoupled from the system. Accordingly, in the incompressible limit there are only four unknown field quantities $(u_{i},p)$ and the momentum balance and continuity equations suffice for the governing field equations.

We pause here to remark that all four field equations \eqref{eq:NS1}, \eqref{eq:continuity1} are \emph{first-order} in time with respect to the field quantities $u_{i}$ and $\rho$. This will be important shortly, when we double the order of the equations. It should also be noted, as mentioned previously, that the first-order problem described above is inherently \emph{non-Hamiltonian}, in that there is no action $\mathcal{S}$ for which Hamilton's principle ($\delta\mathcal{S}=0$) yields the first-order field equations~\cite{Millikan1929,Finlayson1972,Finlayson1972a}. Finally, we note that in the incompressible limit, $\mathcal{R}_{4}$ becomes independent of $\dot{\rho}$ and is no longer first-order in time.

\subsection{Second-order formulation}\label{sec:2ndorder}

Although the first-order formulation of the problem is intrinsically non-Hamiltonian~\cite{Millikan1929,Finlayson1972,Finlayson1972a}, nevertheless a Hamiltonian for the system may be found by considering a second-order formulation. Following Sanders~\cite{Sanders2023b}, we observe that the actual motion of the fluid corresponds to the particular fields $(u_{i},p,\rho)$ for which the following action achieves a local minimum:
\begin{equation}\label{eq:action}
\mathcal{S}^{*}[u_{j},p,\rho]=\int\text{d}^{4}x\left(\frac{1}{2}\mathcal{R}_{i}\mathcal{R}_{i}+\frac{1}{2}\mathcal{R}_{4}\mathcal{R}_{4}\right),
\end{equation}
where $\text{d}^{4}x=\text{d}x_{1}\text{d}x_{2}\text{d}x_{3}\text{d}t$, and the integral is carried out over both the control volume $\mathcal{V}$ occupied by the fluid ($x_{j}\in\mathcal{V}$) and the time period of interest ($t\in[t_{1},t_{2}]$). It must be emphasized that this action contains no new physics. Again, this is simply the principle of least squares~\cite{Finlayson1972a} averaged over the spacetime occupied by the fluid. The entire physics of the problem are already contained in the residuals ($\mathcal{R}_{i}$, $\mathcal{R}_{4}$).

Without an equation of state relating $\rho$ to $p$, the problem is underconstrained with five unknown field quantities and only four dynamical field equations. Anticipating the case of incompressible flow, where the density is constant and the four field quantities are $u_{i}$ and $p$, henceforth we will assume an equation of state of the form $\rho=\hat{\rho}(p)$, with $\hat{\rho}$ a known function determined either from first principles or empirically. In this way, the density field may be eliminated in favor of the pressure field, and the field equations assume the following form:
\begin{equation}\label{eq:NS2}
\mathcal{R}_{i}[u_{j},p;x_{k},t]\equiv \hat{\rho} \dot{u}_{i} + \hat{\rho} u_{i,j}u_{j} + p_{,i} - \mu u_{i,jj} - (\mu+\lambda)u_{j,ji} - \hat{\rho} b_{i} = 0,
\end{equation}
\begin{equation}\label{eq:continuity2}
\mathcal{R}_{4}[u_{j},p]\equiv \hat{\rho}'\dot{p} + \hat{\rho}'p_{,i}u_{i} + \hat{\rho} u_{i,i} = 0,
\end{equation}
where $\hat{\rho}'=\text{d}\hat{\rho}/\text{d}p$. We note that, under equilibrium conditions, $\hat{\rho}'$ is related to the speed of sound $c$ and the bulk modulus $K$ as $\hat{\rho}'=1/c^{2}=\rho/K$ (for incompressible fluids, $\hat{\rho}'\equiv0$ and the speed of sound and bulk modulus are both infinite). Having specified $\hat{\rho}(p)$, $\mu$, $\lambda$, and $b_{i}(x_{j},t)$, and having prescribed appropriate auxiliary conditions (initial and boundary conditions), one seeks the four field quantities $(u_{i},p)$ satisfying the governing field equations and the auxiliary conditions. To recover the case of incompressible flow, we will eventually take $\hat{\rho}'\equiv0$.

We pause here to note that, even though the residuals ($\mathcal{R}_{i}$, $\mathcal{R}_{4}$) vanish for the actual motion, they are \emph{not} trivially zero. That is, the residuals \emph{only} vanish for the \emph{particular} fields $(u_{i},p)$ which satisfy the first-order field equations~\eqref{eq:NS2} and \eqref{eq:continuity2}; they do not vanish for \emph{every conceivable} $(u_{i},p)$. Thus it is not appropriate to take $\mathcal{R}_{i}\equiv0$, $\mathcal{R}_{4}\equiv0$. We will return to this point later when we discuss the Hamiltonian formulation of the problem.

For now, we note that the action $\mathcal{S}^{*}=\mathcal{S}^{*}[u_{i},p]$ defines a Lagrangian
\begin{equation}\label{eq:Lagrangian}
L^{*}[u_{i},p;t]=\int \text{d}^{3}x(\mathcal{L}^{*}) ,
\end{equation}
where the integral is carried out over the volume $\mathcal{V}$ only ($\text{d}^{3}x=\text{d}x_{1}\text{d}x_{2}\text{d}x_{3}$), with Lagrangian density
\begin{equation}\label{eq:Lagrangiandensity}
\mathcal{L}^{*}[u_{j},p;x_{k},t]=\frac{1}{2}\mathcal{R}_{i}\mathcal{R}_{i}+\frac{1}{2}\mathcal{R}_{4}\mathcal{R}_{4}.
\end{equation}
Because the residuals ($\mathcal{R}_{i}$, $\mathcal{R}_{4}$) have been non-dimensionalized, the Lagrangian density is also dimensionless. Once again, even though the Lagrangian vanishes for the actual motion, it is not trivially zero, and it is not appropriate to take $L^{*}\equiv0$.

As noted above, the actual motion of the fluid corresponds to the particular fields $(u_{i},p)$ for which $\mathcal{S}^{*}$ achieves a local minimum. To obtain the Euler-Lagrange equations, the conjugate momenta, and the natural auxiliary conditions, we insist that $\mathcal{S}^{*}$ not vary to first order ($\delta\mathcal{S}^{*}=0$) under small variations in the fields $(\delta u_{i},\delta p)$. Evaluating $\delta\mathcal{S}^{*}$, integrating by parts, and collecting like terms, we find that
\begin{subequations}
\begin{align}
\delta\mathcal{S}^{*}=\int\text{d}^{4}x&\biggl\{\biggl[-\frac{\partial}{\partial t} \left(\hat{\rho}\mathcal{R}_{i}\right) -\frac{\partial}{\partial x_{j}} \left(\hat{\rho} \mathcal{R}_{i}u_{j}\right)+ \hat{\rho} \mathcal{R}_{j}u_{j,i}- \mu \mathcal{R}_{i,jj} \biggr.\biggr. \\
&\biggl.- (\mu+\lambda)\mathcal{R}_{j,ij}+ \hat{\rho}'\mathcal{R}_{4}p_{,i}-\frac{\partial}{\partial x_{i}}\left( \hat{\rho} \mathcal{R}_{4}\right)\biggr]\delta{u}_{i}  \\
+&\biggl[\hat{\rho}'\mathcal{R}_{i} \dot{u}_{i}+ \hat{\rho}' \mathcal{R}_{i}u_{i,j}u_{j}- \mathcal{R}_{i,i}- \hat{\rho}' \mathcal{R}_{i}b_{i}
+ \hat{\rho}''\mathcal{R}_{4}\dot{p}\biggr. \\
&\biggl.\biggl.-\frac{\partial}{\partial t}\left( \hat{\rho}'\mathcal{R}_{4}\right) + \hat{\rho}''\mathcal{R}_{4}p_{,i}u_{i} - \frac{\partial}{\partial x_{i}}\left( \hat{\rho}'\mathcal{R}_{4}u_{i}\right)+ \hat{\rho}' \mathcal{R}_{4}u_{i,i}\biggr] \delta p \biggr\} \\
+\int&\text{d}^{3}x\biggl[\hat{\rho}\mathcal{R}_{i} \delta{u}_{i}+\hat{\rho}'\mathcal{R}_{4}\delta{p}\biggr]_{t_{1}}^{t_{2}} \label{eq:temporalBCs}\\
+\int&\text{d}^{2}x\text{d}t\biggl\{\biggl[\hat{\rho} \mathcal{R}_{i}u_{j}n_{j}+\mu \mathcal{R}_{i,j}n_{j}+ (\mu+\lambda)\mathcal{R}_{j,i}n_{j}+\hat{\rho} \mathcal{R}_{4}n_{i}\biggr]\delta u_{i}\biggr.\\
&\biggl.\biggl[- \mu \mathcal{R}_{i}n_{j}- (\mu+\lambda)\mathcal{R}_{j}n_{i}\biggr]\delta u_{i,j}+\biggl[\mathcal{R}_{i}n_{i}+\hat{\rho}'\mathcal{R}_{4}u_{i}n_{i}\biggr]\delta p\biggr\}, 
\end{align}
\end{subequations}
where the purely volumetric integral ($\text{d}^{3}x$) is carried out over $\mathcal{V}$, the surface integral ($\text{d}^{2}x$) is carried out over the boundary $\partial\mathcal{V}$, and $n_{i}$ is the unit outward normal vector to $\partial\mathcal{V}$. Note that, because we are using Eulerian coordinates $x_{j}$, the volume element $\text{d}^{3}x$ is not to be varied.

The Euler-Lagrange equations (which hold for all $x_{j}\in\mathcal{V}$) may be read directly from the spacetime  ($\text{d}^{4}x$) integral:
\begin{align}
\delta{u}_{i}: \quad &-\frac{\partial}{\partial t} \left(\hat{\rho}\mathcal{R}_{i}\right) -\frac{\partial}{\partial x_{j}} \left(\hat{\rho} \mathcal{R}_{i}u_{j}\right)+ \hat{\rho} \mathcal{R}_{j}u_{j,i}- \mu \mathcal{R}_{i,jj}  \nonumber\\
&- (\mu+\lambda)\mathcal{R}_{j,ij}+ \hat{\rho}'\mathcal{R}_{4}p_{,i}-\frac{\partial}{\partial x_{i}}\left( \hat{\rho} \mathcal{R}_{4}\right)=0, \label{eq:EL1}\\
\delta p:\quad & \biggl.\hat{\rho}'\mathcal{R}_{i} \dot{u}_{i}+ \hat{\rho}' \mathcal{R}_{i}u_{i,j}u_{j}- \mathcal{R}_{i,i}- \hat{\rho}' \mathcal{R}_{i}b_{i}
+ \hat{\rho}''\mathcal{R}_{4}\dot{p}\biggr.\nonumber \\
&-\frac{\partial}{\partial t}\left( \hat{\rho}'\mathcal{R}_{4}\right) + \hat{\rho}''\mathcal{R}_{4}p_{,i}u_{i} - \frac{\partial}{\partial x_{i}}\left(\hat{\rho}'\mathcal{R}_{4}u_{i}\right)+ \hat{\rho}' \mathcal{R}_{4}u_{i,i} =0.\label{eq:EL2}
\end{align}
It should be noted that all four Euler-Lagrange equations \eqref{eq:EL1}, \eqref{eq:EL2} are \emph{second-order} in time, as they involve time derivatives of the residuals. By doubling the order of the equations, we have put the problem in Hamiltonian form, consistent with the general result of Sanders~\cite{Sanders2023b}. We also note that all four Euler-Lagrange equations of the second-order formulation are automatically satisfied by the solution to the first-order formulation (\emph{i.e.}, the actual motion), for which $\mathcal{R}_{i}=0$ and $\mathcal{R}_{4}=0$ everywhere and at all times.

Corresponding to each of the four field quantities is a canonically conjugate ``momentum'' field, which can be read from \eqref{eq:temporalBCs}. The momenta conjugate to the velocities $u_{i}$ are
\begin{equation}
\pi_{i}\equiv \hat{\rho}\mathcal{R}_{i},
\end{equation}
and the momentum conjugate to the pressure $p$ is
\begin{equation}
\pi_{4}\equiv \hat{\rho}'\mathcal{R}_{4}.
\end{equation}
In the forthcoming Hamiltonian formulation, the conjugate momenta will be used to eliminate the (partial) time derivatives $(\dot{u}_{i},\dot{p})$ of the field quantities from the Hamiltonian. In general, Hamilton's principle would insist that the variations $(\delta u_{i},\delta p)$ vanish at the endpoints $t=t_{1}$ and $t=t_{2}$ to ensure that the purely volumetric integral~\eqref{eq:temporalBCs} vanishes identically. Interestingly, for the actual motion ($\mathcal{R}_{i}=0,\mathcal{R}_{4}=0$), the volumetric integral~\eqref{eq:temporalBCs} already vanishes even without taking $(\delta u_{i},\delta p)$ to vanish at $t_{1}$ and $t_{2}$. We interpret this to mean that the actual motion is the natural evolution of the second-order formulation~\cite{Sanders2023b}.

Although the conjugate momenta ($\pi_{i}$, $\pi_{4}$) do not coincide with conventional linear or angular momenta, there is nonetheless a curious mathematical connection between the conjugate momenta and the linear momentum density $P_{i}=\rho u_{i}$, which we will see presently from the natural boundary conditions. These are read directly from the surface ($\text{d}^{2}x$) integral:
\begin{align}
\biggl.\delta u_{i}:\quad &\hat{\rho} \mathcal{R}_{i}u_{j}n_{j}+\mu \mathcal{R}_{i,j}n_{j}+ (\mu+\lambda)\mathcal{R}_{j,i}n_{j}+\hat{\rho} \mathcal{R}_{4}n_{i}=0\biggr.,\label{eq:BC1}\\
\biggl.\delta u_{i,j}:\quad &- \mu \mathcal{R}_{i}n_{j}- (\mu+\lambda)\mathcal{R}_{j}n_{i}=0\biggr.,\label{eq:BC2}\\
\biggl.\delta p:\quad&\mathcal{R}_{i}n_{i}+\hat{\rho}'\mathcal{R}_{4}u_{i}n_{i}=0\biggr.. \label{eq:BC3}
\end{align}
This last condition, \eqref{eq:BC3}, establishes a connection between the new conjugate momenta and the conventional linear momenta. Multiplying \eqref{eq:BC3} by $\hat{\rho}$, and noting that $\hat{\rho}u_{i}=\rho u_{i}= P_{i}$, we find that
\begin{equation}
(\pi_{i}+\pi_{4}P_{i})n_{i}=0.
\end{equation}
Evidently, boundary condition \eqref{eq:BC3} states that the flux of the vector $\Pi_{i}\equiv\pi_{i}+\pi_{4}P_{i}$ through the boundary $\partial\mathcal{V}$ should vanish. It is interesting that this new vector $\Pi_{i}$ contains both old and new momenta, with $\pi_{4}$ ``carried'' (\emph{i.e.}, given direction) by $P_{i}$. The actual physical meaning of these natural boundary conditions is less clear and may require further investigation.

\subsection{Equivalence of the first- and second-order formulations}\label{sec:equivalence}

The first- and second-order formulations are mathematically equivalent, in the sense that imposing \emph{identical auxiliary conditions} on the two formulations will yield identical solutions $(u_{i},p)$. In other words, with identical auxiliary conditions, $(u_{i},p)$ is a solution to the first-order formulation \emph{if and only if} the same $(u_{i},p)$ is a solution to the second-order formulation. 

The proof is straightforward. Consider the two formulations separately, and impose on the second-order formulation identical auxiliary conditions to those of the first-order formulation. In particular, just like the simple example given in Section~\ref{sec:toyproblem}, the second-order formulation requires additional auxiliary conditions over and above those applied to the first-order formulation. These include initial conditions making $\mathcal{R}_{i}(x_{k},0)=0$ and $\mathcal{R}_{4}(x_{k},0)=0$ for all $x_{k}\in\mathcal{V}\cup\partial\mathcal{V}$, along with boundary conditions making $\mathcal{R}_{i}(x_{k},t)=0$, $\mathcal{R}_{i,j}(x_{k},t)=0$, and $\mathcal{R}_{4}(x_{k},t)=0$ for all $x_{k}\in\partial\mathcal{V}$ and all times $t$. By supposition, the auxiliary conditions applied to the two formulations are identical, so it suffices to show that $(u_{i},p)$ satisfies the governing field equations \eqref{eq:NS2}, \eqref{eq:continuity2} of the first-order formulation ($\mathcal{R}_{i}=0$ and $\mathcal{R}_{4}=0$) everywhere in $\mathcal{V}$ and at all times \emph{if and only if} $(u_{i},p)$ satisfies the Euler-Lagrange equations \eqref{eq:EL1}, \eqref{eq:EL2} of the second-order formulation everywhere in $\mathcal{V}$ and at all times.

Suppose first that $(u_{i},p)$ satisfies the governing field equations \eqref{eq:NS2}, \eqref{eq:continuity2} of the first-order formulation everywhere in $\mathcal{V}$ and at all times. Then $\mathcal{R}_{i}=0$, $\mathcal{R}_{4}=0$, and $(u_{i},p)$ is a trivial solution to the Euler-Lagrange equations \eqref{eq:EL1}, \eqref{eq:EL2} of the second-order formulation. 

Conversely, suppose that $(u_{i},p)$ satisfies the Euler-Lagrange equations \eqref{eq:EL1}, \eqref{eq:EL2} of the second-order formulation everywhere in $\mathcal{V}$ and at all times. We note that $(\mathcal{R}_{i}=0,\mathcal{R}_{4}=0)$ constitutes an equilibrium solution of the Euler-Lagrange equations \eqref{eq:EL1}, \eqref{eq:EL2}. Thus, because the initial conditions are chosen such that $\mathcal{R}_{i}(x_{k},0)=0$ and $\mathcal{R}_{4}(x_{k},0)=0$ for all $x_{k}\in\mathcal{V}\cup\partial\mathcal{V}$, and because the boundary conditions are chosen such that $\mathcal{R}_{i}(x_{k},t)=0$, $\mathcal{R}_{i,j}(x_{k},t)=0$,  and $\mathcal{R}_{4}(x_{k},t)=0$ for all $x_{k}\in\partial\mathcal{V}$ and all times $t$, then $\mathcal{R}_{i}$ and $\mathcal{R}_{4}$ will remain identically zero everywhere in $\mathcal{V}$ for all future times. Thus, $(u_{i},p)$ satisfies the governing field equations \eqref{eq:NS2}, \eqref{eq:continuity2} of the first-order formulation everywhere in $\mathcal{V}$ and at all times. This completes the proof, and we have established that the two formulations are equivalent. $\square$

\section{Hamiltonian formulation of the problem}\label{sec:HamiltonianformulationofNSP}

We are now ready to proceed with the Hamiltonian formulation of the problem. The Lagrangian $L^{*}$ has a corresponding Hamiltonian
\begin{equation}\label{eq:Hamiltonian}
H^{*}=\int \text{d}^{3}x(\mathcal{H}^{*}) ,
\end{equation}
with the Hamiltonian density $\mathcal{H}^{*}$ obtained from the Lagrangian density $\mathcal{L}^{*}$ via the Legendre transform
\begin{equation}\label{eq:Hamiltoniandensity}
\mathcal{H}^{*}=\pi_{i}\dot{u}_{i}+\pi_{4}\dot{p}-\mathcal{L}^{*}=\pi_{i}\dot{u}_{i}+\pi_{4}\dot{p}-\frac{1}{2}\mathcal{R}_{i}\mathcal{R}_{i}-\frac{1}{2}\mathcal{R}_{4}\mathcal{R}_{4}.
\end{equation}
Again, this $H^{*}$ has nothing to do with the total mechanical energy of the system, although it \emph{is} a conserved quantity, since $H^{*}=0$ for the actual motion---just as in the example of Section~\ref{sec:toyproblem}. In order to write down Hamilton's equations, we must express $\mathcal{H}^{*}$ in terms of the fields and the conjugate momenta, eliminating $\dot{u}_{i}$ and $\dot{p}$.

Observe that $\mathcal{R}_{i}=\pi_{i}/\hat{\rho}$, and ignoring for the moment the incompressible limit, we may write $\mathcal{R}_{4}=\pi_{4}/\hat{\rho}'$ ($\hat{\rho}'\neq0$). In this way, using the functional expressions for the residuals given by~\eqref{eq:NS2} and~\eqref{eq:continuity2}, we find that
\begin{equation}\label{eq:uidoteqn}
\dot{u}_{i}=\frac{1}{(\hat{\rho})^{2}}{\pi_{i}}-\frac{1}{\hat{\rho}}\biggl(\hat{\rho}u_{i,j}u_{j} +{ p_{,i} - \mu u_{i,jj} - (\mu+\lambda)u_{j,ji}} - \hat{\rho}b_{i}\biggr);
\end{equation}
\begin{equation}\label{eq:pdoteqn}
\dot{p}=\frac{1}{(\hat{\rho}')^{2}}{\pi_{4}}-\frac{1}{\hat{\rho}'}\biggl({\hat{\rho}'p_{,i}u_{i} + \hat{\rho}u_{i,i}}\biggr), \quad \hat{\rho}'\neq0;
\end{equation}
and
\begin{align}
\mathcal{H}^{*}[u_{i},p,\pi_{j},\pi_{4};x_{k},t]=&\frac{1}{2}\frac{1}{(\hat{\rho})^{2}}{\pi_{i}\pi_{i}}-\frac{1}{\hat{\rho}}\biggl(\hat{\rho}u_{i,j}u_{j} +{ p_{,i} - \mu u_{i,jj} - (\mu+\lambda)u_{j,ji}} - \hat{\rho}b_{i}\biggr)\pi_{i} \nonumber\\
&+\frac{1}{2}\frac{1}{(\hat{\rho}')^{2}}{\pi_{4}\pi_{4}}-\frac{1}{\hat{\rho}'}\biggl({\hat{\rho}'p_{,i}u_{i} + \hat{\rho}u_{i,i}}\biggr)\pi_{4}, \quad \hat{\rho}'\neq0. \label{eq:Hamiltoniandensity2}
\end{align}
Hamilton's equations~\cite{Hamilton1834,Hamilton1835} are as follows:
\begin{align}
\dot{u}_{i}&=\frac{\delta H^{*}}{\delta\pi_{i}}, &\dot{p}&=\frac{\delta H^{*}}{\delta\pi_{4}}, \label{eq:Hamilton12} \\
\dot{\pi}_{i}&=-\frac{\delta H^{*}}{\delta u_{i}}, &\dot{\pi}_{4}&=-\frac{\delta H^{*}}{\delta p}, \label{eq:Hamilton34}
\end{align}
where $\delta H^{*}/\delta u_{i}$, $\delta H^{*}/\delta p$, $\delta H^{*}/\delta \pi_{i}$, and $\delta H^{*}/\delta \pi_{4}$ are the Volterra~\cite{Volterra1930} functional derivatives of $H^{*}$ with respect to the field quantities and the conjugate momenta. Equations~\eqref{eq:Hamilton12} reproduce~\eqref{eq:uidoteqn} and~\eqref{eq:pdoteqn}, respectively. Equations~\eqref{eq:Hamilton34} in turn reproduce the Euler-Lagrange equations \eqref{eq:EL1}, \eqref{eq:EL2} of the second-order formulation.

We return now to our previous observation concerning the vanishing of the residuals. While $H^{*}$ vanishes for the \emph{particular} fields $(u_{i},p)$ that satisfy the governing field equations \eqref{eq:NS2}, \eqref{eq:continuity2} of the first-order formulation, it does not vanish for \emph{every conceivable} $(u_{i},p)$. The latter would imply, according to Equations~\eqref{eq:Hamilton12}, that $\dot{u}_{i}\equiv0$ and $\dot{p}\equiv0$, which is not generally the case. This observation, and the fact that Equations~\eqref{eq:Hamilton34} faithfully reproduce the Euler-Lagrange equations \eqref{eq:EL1}, \eqref{eq:EL2} of the second-order formulation, confirm that the Hamiltonian formulation described above is, in fact, a legitimate reformulation of the problem. In the following section, we develop the Hamilton-Jacobi theory as it relates to the present formulation, the goal being to find a canonical transformation to a new set of fields ($\phi_{i}$, $\phi_{4}$) and conjugate momenta for which the Hamiltonian \emph{does} vanish identically.

To obtain the Hamiltonian for incompressible flow, we set $\hat{\rho}'\equiv0$ from the beginning, in which case $\mathcal{R}_{4}$ reduces to $\hat{\rho} u_{i,i}$ and $\pi_{4}$ vanishes identically, consistent with the fact that $\mathcal{R}_{4}$ becomes independent of $\dot{p}$. The Hamiltonian density in turn reduces to
\begin{equation}
\mathcal{H}^{*}=\pi_{i}\dot{u}_{i}-\frac{1}{2}\mathcal{R}_{i}\mathcal{R}_{i}, \quad \hat{\rho}'\equiv0,
\end{equation}
or, in terms of the conjugate momenta,
\begin{equation}
\mathcal{H}^{*}[u_{i},p,\pi_{j};x_{k},t]=\frac{1}{2}\frac{1}{{\rho}^{2}}{\pi_{i}\pi_{i}}-\frac{1}{{\rho}}\biggl({\rho}u_{i,j}u_{j} + p_{,i} - \mu u_{i,jj} - {\rho}b_{i}\biggr)\pi_{i}, \quad \hat{\rho}'\equiv0,\label{eq:Hamiltoniandensity3}
\end{equation} 
where the density $\hat{\rho}=\rho$ is a constant and we have used the fact that $u_{i,i}=0$. Hamilton's equations $\dot{u}_{i}=\delta H^{*}/\delta\pi_{i}$, $\dot{\pi}_{i}=-\delta H^{*}/\delta u_{i}$, and $0\equiv\dot{\pi}_{4}=-\delta H^{*}/\delta p$ still apply (and reproduce the corresponding equations in the incompressible limit), but $\dot{p}=\delta H^{*}/\delta\pi_{4}$ must be replaced by the constraint that $u_{i,i}=0$. That the incompressibility condition should take the place of the pressure equation $\dot{p}=\delta H^{*}/\delta\pi_{4}$ is consistent with the well known result that the pressure usually serves as the Lagrange multiplier for the incompressibility constraint~\cite{Lanczos1986,Badin2018}.

\subsection{Hamilton-Jacobi equation}\label{sec:HamiltonJacobi}

One of the most significant aspects of the Hamiltonian formalism is that it leads to the transformation theory of Hamilton and Jacobi~\cite{Hamilton1833,Hamilton1834,Hamilton1835,Jacobi1837,Jacobi18421843,Whittaker,Lanczos1986}, celebrated both for unifying particle mechanics with wave optics~\cite{Hamilton1833} and for its relationship to the Schr\"{o}dinger equation of quantum mechanics~\cite{Schroedinger1926,Schroedinger1926a}. Here we will obtain a Hamilton-Jacobi equation representing the Navier-Stokes problem.

In the context of discrete mechanics, Hamilton's principal function is obtained as the solution to the Hamilton-Jacobi equation, which is in turn defined by the functional form of the Hamiltonian. Hamilton's principal function provides the generating function for a canonical transformation to a new set of generalized coordinates and conjugate momenta for which the Hamiltonian vanishes identically, in which case Hamilton's equations do, in fact, become trivial. The new coordinates and their conjugate momenta are simply equal to their initial values.

In the present context, the role of Hamilton's principal function is played by a characteristic functional $\text{S}^{*}=\text{S}^{*}[u_{i},p,t]$ (not to be confused with the action $\mathcal{S}^{*}$, although they are related; see Appendix~\ref{app:HJE}), which is the solution to the following Hamilton-Jacobi equation:
\begin{equation}\label{eq:HJE}
{H}^{*}\left[u_{i},p,\frac{\delta \text{S}^{*}}{\delta u_{j}},\frac{\delta \text{S}^{*}}{\delta p};t\right]+\frac{\partial \text{S}^{*}}{\partial t}=0,
\end{equation}
where $\delta \text{S}^{*}/\delta u_{i}$ and $\delta \text{S}^{*}/\delta p$ are the Volterra~\cite{Volterra1930} functional derivatives of $\text{S}^{*}$ with respect to the field quantities. Interested readers will find the derivation of \eqref{eq:HJE} in Appendix~\ref{app:HJE}. Henceforth we will refer to $\text{S}^{*}$ as ``Hamilton's principal functional.'' Substituting for the conjugate momenta in \eqref{eq:Hamiltoniandensity2}, we obtain for the Hamilton-Jacobi equation
\begin{align}
\int \text{d}^{3}x&\left[\frac{1}{2}\frac{1}{(\hat{\rho})^{2}}\frac{\delta \text{S}^{*}}{\delta u_{i}}\frac{\delta \text{S}^{*}}{\delta u_{i}}-\frac{1}{\hat{\rho}}\biggl(\hat{\rho}u_{i,j}u_{j} +{ p_{,i} - \mu u_{i,jj} - (\mu+\lambda)u_{j,ji}} - \hat{\rho}b_{i}\biggr)\frac{\delta \text{S}^{*}}{\delta u_{i}}\right. \nonumber \\
&\left.+\frac{1}{2}\frac{1}{(\hat{\rho}')^{2}}\frac{\delta \text{S}^{*}}{\delta p}\frac{\delta \text{S}^{*}}{\delta p}-\frac{1}{\hat{\rho}'}\biggl({\hat{\rho}'p_{,i}u_{i} + \hat{\rho}u_{i,i}}\biggr)\frac{\delta \text{S}^{*}}{\delta p}\right]+\frac{\partial \text{S}^{*}}{\partial t}=0, \quad \hat{\rho}'\neq0. \label{eq:HamiltonJacobi}
\end{align}
In contrast to the four original field equations---\eqref{eq:NS1} and \eqref{eq:continuity1}---the Hamilton-Jacobi equation~\eqref{eq:HamiltonJacobi} is a \emph{single} equation in Hamilton's principal functional $\text{S}^{*}$. This constitutes an equivalent formulation of the problem, as a complete and nontrivial solution to~\eqref{eq:HamiltonJacobi} is tantamount to an integration of Hamilton's equations \eqref{eq:Hamilton12} and \eqref{eq:Hamilton34} (note that it is not appropriate to take $\text{S}^{*}\equiv0$ for the same reason that it is not appropriate to take $H^{*}\equiv0$). In this way, we have reduced the problem of finding four separate field quantities to that of finding a single functional in those field quantities. One need only deduce (or even guess) the general form of $\text{S}^{*}$ in order to solve the problem. If an analytical expression for $\text{S}^{*}$ can be obtained, it will lead via canonical transformation to a new set of fields ($\phi_{i}$, $\phi_{4}$) and conjugate momenta which are simply equal to their initial values, giving analytical expressions for the four original fields $(u_{i},p)$.

The case of incompressible flow requires care, as $\hat{\rho}'\equiv0$ and $\hat{\rho}'$ appears in the denominators of terms in~\eqref{eq:HamiltonJacobi}. Even so, the Hamiltonian formulation remains well posed in the incompressible limit. Recall that, with $\hat{\rho}'\equiv0$, $\mathcal{R}_{4}$ reduces to $\hat{\rho} u_{i,i}$, $\pi_{4}$ vanishes identically, and the Hamiltonian density reduces to~\eqref{eq:Hamiltoniandensity3}. Substituting for the conjugate momenta $\pi_{i}$ in~\eqref{eq:Hamiltoniandensity3}, the corresponding Hamilton-Jacobi equation is
\begin{equation}\label{eq:HamiltonJacobi2}
\int \text{d}^{3}x\left[\frac{1}{2}\frac{1}{{\rho}^{2}}\frac{\delta \text{S}^{*}}{\delta u_{i}}\frac{\delta \text{S}^{*}}{\delta u_{i}}-\frac{1}{{\rho}}\biggl({\rho}u_{i,j}u_{j} + p_{,i} - \mu u_{i,jj} - {\rho}b_{i}\biggr)\frac{\delta \text{S}^{*}}{\delta u_{i}}\right]+\frac{\partial \text{S}^{*}}{\partial t}=0, \quad \hat{\rho}'\equiv0,
\end{equation}
with $\delta \text{S}^{*}/\delta p=0$, since again $\pi_{4}$ vanishes identically for incompressible flow. Here the merit of starting from the compressible form of the equations becomes fully evident, as it would not necessarily have been clear that $\delta \text{S}^{*}/\delta p$ should vanish in the incompressible limit without knowing that in general $\pi_{4}=\hat{\rho}'\mathcal{R}_{4}$. This is the form of the Hamilton-Jacobi equation as it relates to the traditional Navier-Stokes problem. In this case, the pressure is determined last of all, and is whatever it needs to be to enforce the incompressibility constraint $u_{i,i}=0$ (again consistent with the role of pressure as Lagrange multiplier~\cite{Lanczos1986,Badin2018}). 

It must be acknowledged that the Hamilton-Jacobi equation developed above (either~\eqref{eq:HamiltonJacobi} for the compressible case or~\eqref{eq:HamiltonJacobi2} in the incompressible limit) contains Volterra~\cite{Volterra1930} functional derivatives and is thus by no means trivial to solve. Indeed, it appears that solving such equations is itself a long-standing open problem in mathematics, which has received very little attention since the first half of the twentieth century~\cite{Michal1926,Jordan1928,Levy1951,Tatarskii1961,Syavavko1974,Dieudonne1981,Kovalchik1993}.  Nevertheless, if a rigorous theory of such equations can be developed, the present formulation of the Navier-Stokes problem might be solved as one special case. The present authors submit that such an endeavor is worthwhile and merits further study.

We conclude this section by remarking that, in the inviscid limit ($\mu=\lambda=0$), all of the preceding formalism remains perfectly well-posed. In that limit, the present approach yields a mathematically equivalent second-order formulation of the inviscid Euler equations, as one would expect. Interested readers will find the full details in Appendix~\ref{app:inviscidlimit}.

\section{Discussion}\label{sec:discussion}

In this section we provide some qualitative interpretations of the developments of Section~\ref{sec:LagrangianformulationofNSP}. More specifically, we investigate the incompressible form (via constant, uniform density) of the Euler-Lagrange equations \eqref{eq:EL1} and \eqref{eq:EL2} when the residuals $\mathcal{R}_i$ and $\mathcal{R}_4$ are substituted. 

Our motivation is again the simple example of Section~\ref{sec:toyproblem} for which the first-order non-Hamiltonian system $\dot{v}=-v$ was converted to the second-order Hamiltonian system $\ddot{v}=v$ by (manual) \emph{elimination} of the non-conservative `damping' term $\dot{v}$ (see \cite{Sanders2022} for a similar result for the damped harmonic oscillator converting from second- to fourth-order dynamics). Sanders~\cite{Sanders2023b} showed that the elimination process is `automated' by the definition of the action in the first integral of \eqref{eq:toyaction}, which is generalized to the action in \eqref{eq:action} for our current continuum dynamics problem containing fields.

First consider the pressure equation \eqref{eq:EL2} and corresponding natural boundary condition \eqref{eq:BC3}, which take the following incompressible forms:
\begin{equation}
-\mathcal{R}_{i,i}=0 \quad \forall x_j\in\mathcal{V}, \quad\text{subject to }\quad \mathcal{R}_in_i = 0 \quad \forall x_j\in\partial\mathcal{V}.
\end{equation}
This higher-order field equation is simply the divergence of the residual $\mathcal{R}_i$. Upon substituting for $\mathcal{R}_i$ from \eqref{eq:NS1} and subsequently imposing the incompressible continuity condition $u_{i,i}=\mathcal{R}_4/\rho=0$ from \eqref{eq:continuity1}, we obtain:
\begin{equation}\label{eq:Poisson}
p_{,ii} = -[\rho u_ju_{i,j}]_{,i} + \rho b_{i,i},
\end{equation}
which is a Poisson equation for the pressure. The boundary condition is a Neumann type requiring the specification of the normal pressure gradient, $n_ip_{,i} = p_{,n} \equiv f(x_j,t)$, on the boundary:
\begin{equation}\label{eq:pBC}
f(x_j,t) = -n_i\big[\rho\dot{u}_i+\rho u_ju_{i,j}-\mu u_{i,jj}-\rho b_i\big].
\end{equation}
Equation~\eqref{eq:Poisson} and boundary condition \eqref{eq:pBC} evolve the pressure in a manner that ensures the velocity field is solenoidal. This is a well known pressure-velocity based formulation commonly used in the numerical solution of incompressible flows (\textit{e.g.} \cite{Ferziger2002,Pozrikidis2009}).

Next, we consider the velocity equations \eqref{eq:EL1} which, at present, have a more elusive physical interpretation. Here, we instead begin with the natural boundary conditions \eqref{eq:BC1} and \eqref{eq:BC2}, which are due to the $\delta u_i$ and $\delta u_{i,j}$ variations. The incompressible versions of these equations are:
\begin{eqnarray}
\rho \mathcal{R}_{i}u_{j}n_{j}+\mu \mathcal{R}_{i,j}n_{j}=0 \quad\text{and}\quad - \mu \mathcal{R}_{i}n_{j}=0 \quad \forall x_j\in\partial\mathcal{V}.
\end{eqnarray}
The boundary conditions involving the residual $\mathcal{R}_i$ are those compatible with the first-order Navier-Stokes equations, such as the no-slip and no-penetration conditions. Indeed, if we specify the velocity vector of the \emph{actual motion} on the boundary, then $\mathcal{R}_i\equiv 0$ there. Note that the pressure of the actual motion on the boundary will be known from the simultaneous solution of \eqref{eq:Poisson}.

However, the \emph{gradient} terms $\mathcal{R}_{i,j}$ will introduce up to third-order spatial derivatives that must be specified. These represent the additional boundary conditions that must accompany the higher-order governing equation, which will be seen shortly to be second-order in time and fourth-order in space. Again, recall the example of Section~\ref{sec:intro}, whereby the system \eqref{eq:toy2} must be appended with a second (initial) condition specifying the (time) derivative of the coordinate $v(t)$. In the present context, these boundary conditions are ostensibly tantamount to specification of the viscous stress on the boundary by way of velocity gradients. 

In general, the conditions at a boundary require two \textit{transition relations} \cite{Batchelor2000,White2006} to ultimately describe the momentum transport. Mathematically speaking, these conditions are the jump in velocity (momentum intensity) and the jump in stress (momentum flux). Under ordinary physical circumstances the velocity and stress are assumed to be continuous. However, this is one particular form of the transition relations, and there are familiar examples to which they do not apply. For example, at a liquid-gas interface the stress relation is modified to account for a non-zero jump in the normal stress that is balanced by a force due to surface tension (the tangential stress component usually still taken to be continuous). Similarly, in the event that molecular slip occurs, the typical transition relation gives an expression for the slip velocity (\textit{e.g.} \cite{Troian1997,Thalakkottor2016}). In the case of energy transport, analogous conditions are needed regarding jumps in temperature (intensity) and heat flow (flux), which are recognized as the concept of thermal contact resistance.

We now turn our attention to the Euler-Lagrange equations \eqref{eq:EL1}, which upon imposing incompressibility and expanding derivatives of product terms yields:
\begin{eqnarray}\label{eq:EL2a}
\rho\dot{\mathcal{R}}_i + \rho u_j\mathcal{R}_{i,j} = \rho \mathcal{R}_{j}u_{j,i}- \mu \mathcal{R}_{i,jj} \quad \forall x_j\in\mathcal{V}.
\end{eqnarray}
The left-hand side is the material derivative of the residual $\mathcal{R}_i$. Our purpose here is to observe which terms from the first-order Navier-Stokes equation are `eliminated' in the higher-order formulation. Specifically, we are interested in the non-conservative viscous terms; while the body force $b_i$ could also be non-conservative, we will not concern ourselves with this possibility. Direct substitution of $\mathcal{R}_i$ into \eqref{eq:EL2a} generates many terms, but it is found that only one is canceled: the viscous Laplacian of the (time derivative of the) velocity, namely $\mu \dot{u}_{i,jj}$. This term mutually appears from the $\rho\dot{\mathcal{R}}_i$ and $-\mu\mathcal{R}_{i,jj}$ terms in \eqref{eq:EL2a}. To maintain notional clarity, we write the residual as:
\begin{equation}
\mathcal{R}_i = \rho\dot{u}_i - \mu u_{i,kk} + \tilde{\mathcal{R}}_i,
\end{equation}
where index $k$ has been used to avoid confusion with gradient operators in \eqref{eq:EL2a} having index $j$, and $\tilde{\mathcal{R}}_i = \rho u_ku_{i,k}-\rho b_i$ are the remaining terms in the residual. Substituting the above into the first and last terms of \eqref{eq:EL2a}, canceling the aforementioned $\mu \dot{u}_{i,jj}$ term, and then dividing out by the density gives:
\begin{equation}\label{eq:EL2b}
\rho\ddot{u}_i+\dot{\tilde{\mathcal{R}}}_i+u_j\mathcal{R}_{i,j} = \mathcal{R}_ju_{j,i} - \nu\big[-\mu u_{i,kkjj} + \tilde{\mathcal{R}}_{i,jj} \big],
\end{equation}
where $\nu=\mu/\rho$ is the kinematic viscosity (recall that all variables are non-dimensional). We see that this equation is second-order in time and fourth-order in space. Viscous terms still appear in the equation including second- and third-order spatial derivatives. Nevertheless, the technique detailed by Sanders~\cite{Sanders2023b} and employed here evidently ensures that \eqref{eq:EL2b} has a corresponding Hamiltonian structure.

\section{Case study}\label{sec:examples}

We can explore how this method can be applied by considering a simplified example with a known field solution. In looking at the variety of cases in which the Navier-Stokes equations have a known analytical solution, the simplest are those involving steady flows. While the Euler-Lagrange equations~\eqref{eq:EL1}, \eqref{eq:EL2} can be written for these cases, the corresponding Hamilton-Jacobi equation is trivial because for steady flows the fields are already equal to their initial values.

It is therefore worthwhile to examine the simplest unsteady flows, which should result in a nontrivial Hamilton-Jacobi equation. Indeed, there exists a class of flows for which the Navier-Stokes equations take the same simplified form: those in which the flow is incompressible and unidirectional~\cite{Batchelor2000}. This class of problems include both of Stokes's flows~\cite{Stokes1851}, in which a semi-infinite fluid is influenced by a boundary moving in its own plane. In the first of these cases, the boundary is impulsively started and in the second, the boundary oscillates. We can also include developing flow in a channel or pipe. The only difference between these flows results from initial and boundary conditions, but the Navier-Stokes equations and therefore the present Hamilton-Jacobi equation take the same form.

Here we will examine the case in which there is motion only in the $x_1$ direction, and the velocities take the form $\{u_{i}\} = \{u_{1}(x_{2},t),0,0\}$. In the absence of a body force, our pressure gradient in the $x_1$ direction is solely a function of time and the pressure gradients in the $x_2$ and $x_3$ directions are zero. There are thus only two unknown field quantities: $u_{1}(x_{2},t)$ and $p(x_{1},t)$, where $p$ is linear in $x_{1}$. The field equation of primary interest is
\begin{equation}\label{eq:ex1-NS1}
\mathcal{R}_{1} \equiv \rho \dot{u}_{1} + p_{,1} - \mu u_{1,22} = 0,
\end{equation}
and the remaining field equations are satisfied automatically by the assumed form of the fields. Following the procedure described above, the momenta conjugate to $u_{1}$ and $p$ are given by
\begin{equation}
\pi_{1}\equiv \rho\mathcal{R}_{1}, \quad \pi_{4}\equiv0.\label{eq:ex1-pi1}
\end{equation}
This results in a Hamiltonian density given by:
\begin{equation}
\mathcal{H}^{*}=\frac{1}{2}\frac{1}{{\rho}^{2}}\pi_{1}\pi_{1}-\frac{1}{{\rho}}\biggl(p_{,1}-\mu u_{1,22}\biggr)\pi_{1}.
\end{equation}
Hamilton's principal functional $\text{S}^{*}=\text{S}^{*}[u_{1},p,t]$ can be expressed as an integral over $x_{2}$ only, since the other spatial coordinates do not appear and may be integrated out. In this way, we may write the Hamilton-Jacobi equation as follows:
\begin{equation}
\int \text{d}x_2\left[\frac{1}{2}\frac{1}{{\rho}^{2}}\frac{\delta \text{S}^{*}}{\delta u_{1}}\frac{\delta \text{S}^{*}}{\delta u_{1}}-\frac{1}{{\rho}}\biggl(p_{,1}-\mu u_{1,22}\biggr)\frac{\delta \text{S}^{*}}{\delta u_{1}}\right]+\frac{\partial \text{S}^{*}}{\partial t}=0, \label{eq:ex1-HamiltonJacobi2}
\end{equation}
with $\delta\text{S}^{*}/\delta p=0$. The solution to \eqref{eq:ex1-HamiltonJacobi2} would provide a canonical transformation to a new set of coordinates, giving analytical expressions for $(u_{1},p)$. 

Despite knowing the analytical solution for these fields in this particular example, the present authors have not been able to solve this Hamilton-Jacobi equation, since again the solution of such equations is itself an open problem~\cite{Michal1926,Jordan1928,Levy1951,Tatarskii1961,Syavavko1974,Dieudonne1981,Kovalchik1993}. This example therefore appears to be a good place to start for tackling the general problem. Another interesting example to consider might be a 2D Taylor-Green vortex such as that considered by Wu~\emph{et al.}~\cite{Wu2006}.

\section{Conclusion}\label{sec:conclusion}

This paper has presented a novel Hamiltonian formulation of the isotropic Navier-Stokes problem for both compressible and incompressible fluids. This canonical formulation opens several previously unexplored avenues toward a final resolution of the problem, which we briefly describe below.

Perhaps the most obvious route would be to solve the Hamilton-Jacobi equation---either~\eqref{eq:HamiltonJacobi} for the compressible case or~\eqref{eq:HamiltonJacobi2} for the incompressible case---for Hamilton's principal functional $\text{S}^{*}[u_{i},p,t]$ directly. If a complete solution for $\text{S}^{*}$ can be found, it will lead via canonical transformation to a new set of fields which are equal to their initial values, thereby giving analytical expressions for the original velocity and pressure fields. Alternatively, if one can simply establish based on emerging analytical techniques that a complete solution to this Hamilton-Jacobi equation does (or does not) exist under the usual assumptions, that will also settle the question of existence of solutions.

An alternative strategy might be to investigate the corresponding Lagrangian formulation based on the action $\mathcal{S}^{*}$ as given by~\eqref{eq:action}. Because the first- and second-order formulations are mathematically equivalent (recall the proof in Section~\ref{sec:equivalence}), $\mathcal{S}^{*}$ must have as many local minima as there are solutions to the traditional, first-order formulation. Intuitively, it seems as though it ought to be possible to determine---or at least to establish bounds on---the number of critical points an action has based on the form of the Lagrangian~\cite{VandenBerg2002,Kalies2004}. If one can establish that, under the usual assumptions, $\mathcal{S}^{*}$ always has exactly one local minimum, or else demonstrate that there are cases where it fails to achieve a local minimum, that too will resolve the question of existence and uniqueness.

By no means is either of the above programs trivial. As pointed out in Section~\ref{sec:HamiltonJacobi}, solving equations containing Volterra~\cite{Volterra1930} functional derivatives is itself a long-standing open problem in mathematics, which has received very little attention since the first half of the twentieth century~\cite{Michal1926,Jordan1928,Levy1951,Tatarskii1961,Syavavko1974,Dieudonne1981,Kovalchik1993}.  One might even go so far as to call it a ``forgotten'' open problem (as did one of the reviewers of the present paper, who generously drew our attention to References~\cite{Jordan1928,Levy1951,Tatarskii1961,Syavavko1974,Dieudonne1981,Kovalchik1993}). We see the lack of work on such equations as a challenge, yes, but at the same time we also see it as a significant opportunity for advancing the field of analytical continuum mechanics. Perhaps, despite an apparent increase in complexity, a rigorous theory of such equations can be developed after all, in which case the present formulation of the Navier-Stokes problem might be solved as one example. We submit that, at the very least, such an endeavor merits further study, which we intend to continue in future work.

Finally, it is worth noting that the techniques employed here are by no means specific to the Navier-Stokes problem, nor are they restricted to the field of classical mechanics. The suitably-averaged principle of least squares~\cite{Sanders2021,Sanders2022,Sanders2023a,Sanders2023,Sanders2023b} can be applied to any traditionally non-Hamiltonian dynamical system in order to formulate a mathematically equivalent higher-order Hamiltonian system. It is believed that this fundamental result will also find uses in other branches of pure and applied mathematics.

\section*{Acknowledgments}

The authors wish to thank Maggie Sanders for posing thought-provoking questions during the development of this paper. 

We would also like to thank the reviewers for taking the time to provide genuinely constructive feedback and suggestions, which have improved the quality of the paper immeasurably. One reviewer in particular provided not only a number of important technical clarifications, but also a list of 35 additional references to augment the literature review. We express our profound gratitude for the reviewers' efforts.

\section*{Statements and declarations}

\subsection*{Funding}

The authors declare that no funds, grants, or other support were received during the preparation of this manuscript.

\subsection*{Declaration of interests}

The authors report no conflict of interest.

\subsection*{Data availability}

Any data appearing in the present work are available from the corresponding author upon reasonable request.

\subsection*{Author ORCID}

\begin{itemize}[noitemsep]
\item J. W. Sanders, https://orcid.org/0000-0003-3059-3815
\item A. C. DeVoria, https://orcid.org/0000-0001-5615-0807
\item N. J. Washuta, https://orcid.org/0000-0002-4575-0564
\item G. A. Elamin, https://orcid.org/0009-0009-1055-9688
\item K. L. Skenes, https://orcid.org/0000-0002-8371-5006
\item J. C. Berlinghieri, https://orcid.org/0000-0001-7921-1895
\end{itemize}

\subsection*{Author contributions}

Sanders conceived the original idea for the work, wrote the Abstract, Introduction, Lagrangian formulation, Hamiltonian formulation, Conclusion, and Appendix, and contributed to the Literature Review. DeVoria wrote the Discussion, contributed to the Analysis and Literature Review, and provided expertise in the area of fluid mechanics. Washuta wrote the Case Study, contributed to the Literature Review, and provided expertise in the area of fluid mechanics. Elamin contributed to the Literature Review and provided expertise in the area of fluid mechanics. Skenes organized and wrote the Literature Review and provided expertise in the area of fluid mechanics. Berlinghieri contributed to the Introduction and Literature Review, and provided expertise in the areas of analytical mechanics and variational calculus. All authors contributed equally in checking the accuracy of the work, discussing the results together as a group on multiple occasions, drawing conclusions, and proofreading the manuscript.

\subsection*{Ethical guidelines}

Not applicable.

\titleformat{\section}{\bfseries}{Appendix~\thesection .}{1em}{}

\begin{appendices}

\numberwithin{equation}{section}

\section{Derivation of the Hamilton-Jacobi equation}\label{app:HJE}

In what follows, it is important to distinguish between two sets of solutions: solutions to the second-order Euler-Lagrange equations \eqref{eq:EL1} and \eqref{eq:EL2}, and solutions to the first-order field equations \eqref{eq:NS2} and \eqref{eq:continuity2}. The latter are a subset of the former, but not vice versa.

We define \emph{Hamilton's principal functional} $\text{S}^{*}[u_{i},p,t]$ as
\begin{equation}\label{eq:HFdef}
\text{S}^{*}[u_{i},p,t]\equiv\int_{t_{0}}^{t}\text{d}t\left(\tilde{L}^{*}\right),
\end{equation}
where $t$ is the current (variable) time, $t_{0}$ is an arbitrary initial time, and $\tilde{L}^{*}$ denotes the Lagrangian \eqref{eq:Lagrangian} evaluated for fields satisfying the \emph{second-order} Euler-Lagrange equations \eqref{eq:EL1} and \eqref{eq:EL2}---not necessarily the first-order field equations \eqref{eq:NS2} and \eqref{eq:continuity2}. Crucially, because the fields do not necessarily satisfy the first-order field equations, it is not appropriate to take $\text{S}^{*}\equiv0$. We imagine that the time integral has already been carried out, so that the functional $\text{S}^{*}$ may be regarded as an integral over $\mathcal{V}$, that is,
\begin{equation}
\text{S}^{*}[u_{i},p,t]=\int\text{d}^{3}x\left(s^{*}\right),
\end{equation}
where $s^{*}$ is the Lagrangian density \eqref{eq:Lagrangiandensity} evaluated for fields satisfying the {second-order} Euler-Lagrange equations (which will be denoted $\tilde{\mathcal{L}}^{*}$) integrated over time from $t_{0}$ to $t$. 

Starting from \eqref{eq:HFdef} and evaluating the first variation of $\text{S}^{*}$ as we did with the action $\mathcal{S}^{*}$ in Section~\ref{sec:LagrangianformulationofNSP}, we find that
\begin{equation}
\delta\text{S}^{*}=\int\text{d}^{3}x\biggl[\pi_{i}\delta{u}_{i}+\pi_{4}\delta{p}\biggr]_{t_{0}}^{t},
\end{equation}
where we have used the fact that the second-order Euler-Lagrange equations \eqref{eq:EL1} and \eqref{eq:EL2} are satisfied by definition and we have also enforced the natural boundary conditions. But, by the definition of Volterra~\cite{Volterra1930} functional derivatives, we have that
\begin{equation}
\delta\text{S}^{*}=\int\text{d}^{3}x\biggl[\frac{\delta\text{S}^{*}}{\delta u_{i}}\delta{u}_{i}+\frac{\delta\text{S}^{*}}{\delta p}\delta{p}\biggr].
\end{equation}
In this way, we may identify the conjugate momenta with the functional derivatives of $\text{S}^{*}$:
\begin{equation}\label{eq:momentaasfunctionalderivatives}
\pi_{i}=\frac{\delta\text{S}^{*}}{\delta u_{i}}, \quad \pi_{4}=\frac{\delta\text{S}^{*}}{\delta p}.
\end{equation}

Now, starting from \eqref{eq:HFdef} and evaluating the time derivative, we find that
\begin{equation}
\frac{\text{d}\text{S}^{*}}{\text{d}t}=\tilde{L}^{*}=\int\text{d}^{3}x\left(\tilde{\mathcal{L}}^{*}\right).
\end{equation}
But, by the chain rule,
\begin{equation}
\frac{\text{d}\text{S}^{*}}{\text{d}t}=\frac{\partial\text{S}^{*}}{\partial t}+\int\text{d}^{3}x\biggl[\frac{\delta\text{S}^{*}}{\delta u_{i}}\dot{u}_{i}+\frac{\delta\text{S}^{*}}{\delta p}\dot{p}\biggr].
\end{equation}
In this way, we find that
\begin{equation}
\int\text{d}^{3}x\biggl[\frac{\delta\text{S}^{*}}{\delta u_{i}}\dot{u}_{i}+\frac{\delta\text{S}^{*}}{\delta p}\dot{p}-\tilde{\mathcal{L}}^{*}\biggr]+\frac{\partial\text{S}^{*}}{\partial t}=0.
\end{equation}
Here the integral is simply the Hamiltonian $H^{*}$, with the conjugate momenta replaced by the functional derivatives in accordance with~\eqref{eq:momentaasfunctionalderivatives}. Hence we arrive at the Hamilton-Jacobi equation
\begin{equation}
{H}^{*}\left[u_{i},p,\frac{\delta \text{S}^{*}}{\delta u_{j}},\frac{\delta \text{S}^{*}}{\delta p};t\right]+\frac{\partial \text{S}^{*}}{\partial t}=0,
\end{equation}
as claimed in~\eqref{eq:HJE}.

\section{Inviscid flow}\label{app:inviscidlimit}

In this appendix, we consider what becomes of the present formulation in the inviscid limit, \emph{i.e.}, upon setting the viscosities $\mu=\lambda=0$. While this is a relatively simple exercise, it is of interest by virtue of its relationship to the classical Euler equations, which of course are already conservative and so do not require a higher-order formulation to put them in Hamiltonian form~\cite{Olver1982}.

The vanishing of viscous forces affects the residuals of the momentum balance equations:
\begin{equation}\label{eq:NS1inviscid}
\overline{\mathcal{R}}_{i}[u_{j},p,\rho;x_{k},t]\equiv \rho \dot{u}_{i} + \rho u_{i,j}u_{j} + p_{,i} - \rho b_{i} = 0,
\end{equation}
where we are using a bar to distinguish the \emph{inviscid} residuals ($\overline{\mathcal{R}}_{i}$) from their more general forms given in~\eqref{eq:NS1}. Applying the equation of state $\rho=\hat{\rho}(p)$, we obtain
\begin{equation}\label{eq:NS2inviscid}
\overline{\mathcal{R}}_{i}[u_{j},p;x_{k},t]\equiv \hat{\rho} \dot{u}_{i} + \hat{\rho} u_{i,j}u_{j} + p_{,i} - \hat{\rho} b_{i} = 0.
\end{equation}
The residual $\mathcal{R}_{4}$ of the mass balance is unaffected by viscous effects and will remain as it was in the main text.

The Lagrangian density is now
\begin{equation}\label{eq:Lagrangiandensityinviscid}
\overline{\mathcal{L}}^{*}[u_{j},p;x_{k},t]=\frac{1}{2}\overline{\mathcal{R}}_{i}\overline{\mathcal{R}}_{i}+\frac{1}{2}\mathcal{R}_{4}\mathcal{R}_{4},
\end{equation}
and the associated Euler-Lagrange equations are as follows:
\begin{align}
\delta{u}_{i}: \quad &-\frac{\partial}{\partial t} \left(\hat{\rho}\overline{\mathcal{R}}_{i}\right) -\frac{\partial}{\partial x_{j}} \left(\hat{\rho} \overline{\mathcal{R}}_{i}u_{j}\right)+ \hat{\rho} \overline{\mathcal{R}}_{j}u_{j,i} + \hat{\rho}'\mathcal{R}_{4}p_{,i}-\frac{\partial}{\partial x_{i}}\left( \hat{\rho} \mathcal{R}_{4}\right)=0, \label{eq:EL1inviscid}\\
\delta p:\quad & \biggl.\hat{\rho}'\overline{\mathcal{R}}_{i} \dot{u}_{i}+ \hat{\rho}' \overline{\mathcal{R}}_{i}u_{i,j}u_{j}- \overline{\mathcal{R}}_{i,i}- \hat{\rho}' \overline{\mathcal{R}}_{i}b_{i}
+ \hat{\rho}''\mathcal{R}_{4}\dot{p}\biggr.\nonumber \\
&-\frac{\partial}{\partial t}\left( \hat{\rho}'\mathcal{R}_{4}\right) + \hat{\rho}''\mathcal{R}_{4}p_{,i}u_{i} - \frac{\partial}{\partial x_{i}}\left(\hat{\rho}'\mathcal{R}_{4}u_{i}\right)+ \hat{\rho}' \mathcal{R}_{4}u_{i,i} =0.\label{eq:EL2inviscid}
\end{align}
Again, all four Euler-Lagrange equations are {second-order} in time, as they involve time derivatives of the residuals. This is a \emph{mathematically equivalent second-order formulation of the inviscid Euler equations}, as claimed in the main text.

The momenta conjugate to the velocities $u_{i}$ are now
\begin{equation}
\overline{\pi}_{i}\equiv \hat{\rho}\overline{\mathcal{R}}_{i},
\end{equation}
and the momentum $\pi_{4}$ conjugate to the pressure $p$ remains as it was in the main text. For compressible flows, the resulting Hamiltonian density takes the form
\begin{align}
\overline{\mathcal{H}}^{*}[u_{i},p,\overline{\pi}_{j},\pi_{4};x_{k},t]=&\frac{1}{2}\frac{1}{(\hat{\rho})^{2}}{\overline{\pi}_{i}\overline{\pi}_{i}}-\frac{1}{\hat{\rho}}\biggl(\hat{\rho}u_{i,j}u_{j} + p_{,i} - \hat{\rho}b_{i}\biggr)\overline{\pi}_{i} \nonumber\\
&+\frac{1}{2}\frac{1}{(\hat{\rho}')^{2}}{\pi_{4}\pi_{4}}-\frac{1}{\hat{\rho}'}\biggl({\hat{\rho}'p_{,i}u_{i} + \hat{\rho}u_{i,i}}\biggr)\pi_{4}, \quad \hat{\rho}'\neq0. \label{eq:Hamiltoniandensity2inviscid}
\end{align}
Once again, the higher-order Hamiltonian has nothing to do with the total mechanical energy of the system, but rather vanishes for the actual motion (just as it does in the case of viscous flows). With this Hamiltonian density, Hamilton's equations assume the usual canonical form:
\begin{align}
\dot{u}_{i}&=\frac{\delta \overline{H}^{*}}{\delta\overline{\pi}_{i}}, &\dot{p}&=\frac{\delta \overline{H}^{*}}{\delta\pi_{4}}, \label{eq:Hamilton12inviscid} \\
\dot{\overline{\pi}}_{i}&=-\frac{\delta \overline{H}^{*}}{\delta u_{i}}, &\dot{\pi}_{4}&=-\frac{\delta \overline{H}^{*}}{\delta p}, \label{eq:Hamilton34inviscid}
\end{align}
where $\overline{H}^{*}$ denotes $\overline{\mathcal{H}}^{*}$ integrated over the volume $\mathcal{V}$. Equations~\eqref{eq:Hamilton34inviscid} reproduce the Euler-Lagrange equations \eqref{eq:EL1inviscid} and \eqref{eq:EL2inviscid}. In the case of incompressible flows, the Hamiltonian density reduces to
\begin{equation}
\overline{\mathcal{H}}^{*}[u_{i},p,\overline{\pi}_{j};x_{k},t]=\frac{1}{2}\frac{1}{{\rho}^{2}}{\overline{\pi}_{i}\overline{\pi}_{i}}-\frac{1}{{\rho}}\biggl({\rho}u_{i,j}u_{j} + p_{,i} - {\rho}b_{i}\biggr)\overline{\pi}_{i}, \quad \hat{\rho}'\equiv0.\label{eq:Hamiltoniandensity3inviscid}
\end{equation} 
Hamilton's equations $\dot{u}_{i}=\delta \overline{H}^{*}/\delta\overline{\pi}_{i}$, $\dot{\overline{\pi}}_{i}=-\delta \overline{H}^{*}/\delta u_{i}$, and $0\equiv\dot{\pi}_{4}=-\delta \overline{H}^{*}/\delta p$ still apply, but $\dot{p}=\delta \overline{H}^{*}/\delta\pi_{4}$ must be replaced by the constraint that $u_{i,i}=0$, once again consistent with the fact that the pressure serves as Lagrange multiplier for the incompressibility constraint~\cite{Lanczos1986,Badin2018}.

The Hamilton-Jacobi equation
\begin{equation}\label{eq:HJEinviscid}
\overline{H}^{*}\left[u_{i},p,\frac{\delta \overline{\text{S}}^{*}}{\delta u_{j}},\frac{\delta \overline{\text{S}}^{*}}{\delta p};t\right]+\frac{\partial \overline{\text{S}}^{*}}{\partial t}=0
\end{equation}
assumes the following forms for compressible and incompressible flows, respectively:
\begin{align}
\int \text{d}^{3}x&\left[\frac{1}{2}\frac{1}{(\hat{\rho})^{2}}\frac{\delta \overline{\text{S}}^{*}}{\delta u_{i}}\frac{\delta \overline{\text{S}}^{*}}{\delta u_{i}}-\frac{1}{\hat{\rho}}\biggl(\hat{\rho}u_{i,j}u_{j} + p_{,i} - \hat{\rho}b_{i}\biggr)\frac{\delta \overline{\text{S}}^{*}}{\delta u_{i}}\right. \nonumber \\
&\left.+\frac{1}{2}\frac{1}{(\hat{\rho}')^{2}}\frac{\delta \overline{\text{S}}^{*}}{\delta p}\frac{\delta \overline{\text{S}}^{*}}{\delta p}-\frac{1}{\hat{\rho}'}\biggl({\hat{\rho}'p_{,i}u_{i} + \hat{\rho}u_{i,i}}\biggr)\frac{\delta \overline{\text{S}}^{*}}{\delta p}\right]+\frac{\partial \overline{\text{S}}^{*}}{\partial t}=0, \quad \hat{\rho}'\neq0, \label{eq:HamiltonJacobiinviscid}
\end{align}
and
\begin{equation}\label{eq:HamiltonJacobi2inviscid}
\int \text{d}^{3}x\left[\frac{1}{2}\frac{1}{{\rho}^{2}}\frac{\delta \overline{\text{S}}^{*}}{\delta u_{i}}\frac{\delta \overline{\text{S}}^{*}}{\delta u_{i}}-\frac{1}{{\rho}}\biggl({\rho}u_{i,j}u_{j} + p_{,i} - {\rho}b_{i}\biggr)\frac{\delta \overline{\text{S}}^{*}}{\delta u_{i}}\right]+\frac{\partial \overline{\text{S}}^{*}}{\partial t}=0, \quad \hat{\rho}'\equiv0,
\end{equation}
where in the latter case we have $\delta \overline{\text{S}}^{*}/\delta p=0$, since $\pi_{4}$ vanishes identically for incompressible flow.

The second-order formulation described above is fundamentally different from---though again still mathematically equivalent to---the classical Hamiltonian formulation of the first-order Euler equations~\cite{Olver1982}.

\end{appendices}

\clearpage

\begin{spacing}{0.25}
\small
\bibliographystyle{asmems4}
\bibliography{NSTF2023_revised}
\end{spacing}

\end{document}